\RequirePackage{fix-cm}
\documentclass[twocolumn]{svjour3}       
\smartqed  
\usepackage{graphicx}
\usepackage{tabularx}
\usepackage{placeins}
\begin{document}

\title{\LARGE\bf Dissecting the Bond Formation Process of $d^{10}$-Metal-Ethene Complexes with Multireference Approaches}


\author{Yilin Zhao$^{\rm 1}$\and  Katharina Boguslawski$^{\rm 1}$ \and Pawe{\l} Tecmer$^{\rm 1}$ \and Corinne Duperrouzel$^{\rm 1}$\and Gergely Barcza$^{\rm 2}$\and \"Ors Legeza$^{\rm 2}$\and and Paul W. Ayers$^{\rm 1}$
}


\institute{1 \at
              Department of Chemistry and Chemical Biology, McMaster University, Hamilton, 1280 Main Street West, L8S 4M1, Canada\\
              Tel.: +905-525-9140\\
              Fax: +905-522-2509\\
              \email{bogusl@mcmaster.ca}\\
              \email{tecmer@mcmaster.ca}
           \and
           2 \at
              Strongly Correlated Systems ``Lend\"ulet" Research Group, Wigner Research Center for Physics, H-1525 Budapest, Hungary\\
              \email{legeza.ors@wigner.mta.hu}
}

\date{Received: date / Accepted: date}

\maketitle

\begin{abstract}
The bonding mechanism of ethene to a nickel or palladium center is studied by the density matrix renormalization group algorithm, the complete active space self consistent field method, coupled cluster theory, and density functional theory. 
Specifically, we focus on the interaction between the metal atom and bis-ethene ligands in perpendicular and parallel orientations.  
The bonding situation in these structural isomers is further scrutinized using energy decomposition analysis and quantum information theory. 
Our study highlights the fact that when two ethene ligands are oriented perpendicular to each other, the complex is stabilized by the metal-to-ligand double-back-bonding mechanism. Moreover, we demonstrate that nickel-ethene complexes feature a stronger and more covalent interaction between the ligands and the metal center than palladium-ethene compounds with similar coordination spheres.

\end{abstract}

\section{Introduction}\label{intro}
$d^{10}$-transition metals like Ni, Pd, and Pt are very versatile metals used in batteries, alloys, and catalysts. In particular in organometallic chemistry, 
they play an important role in catalytic processes like coupling reactions~\cite{Milstein1979,Denmark2003,Sindlinger2015,Zeni2004} and
cycloaddition reactions~\cite{Nishimura2012,Trost2012}. Unsaturated organic compounds like olefins can easily
form organometallic complexes with $d^{10}$-transition metals by metal--olefin bonding. It is commonly believed that the so-called metal--olefin bonds are formed by the process of back-donation. 

The Dewar-Chatt-Duncanson model~\cite{chatt1953586,dewar1951description} is widely used to explain the back-donation process between metals and olefins. Metal $d$-orbitals overlap with
olefin $\pi^*$-orbitals, allowing electron transfer from metal $d$-orbitals to ligand $\pi^*$-orbitals. The electron transfer process 
from bonding metal to anti-bonding ligand orbitals reduces the bond order of the ligand $\pi$-bonds. This destabilizes the carbon--carbon double bond and lowers the energy barrier to bond cleavage~\cite{Palladium-cat-rev,Chem-Rev-2000,Pd-Chem-Rev-2000,Ozin1977,Jarque1987,Blomberg-1992,Minaev1998}.
\begin{figure}[tb]
\centering
\includegraphics[width=0.4\textwidth]{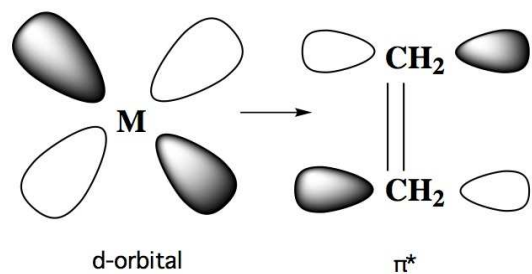}
\caption{Schematic representation of the back-donation process. M refers to either palladium or nickel.}
\label{fig:1}     
\end{figure}

Recently, we elucidated the nickel-ethene reaction pathway and the crucial role of metal-to-ligand back-donation in
the metal-olefin bond-formation process~\cite{Duperrouzel2015}. Our study reveals the presence of a transition state along the nickel-ethene reaction pathway. This peculiar feature in the metal--olefin bond formation process motivates this study of other $d^{10}$-transition metal complexes containing a nickel and palladium centers and their reactions with small olefin ligands~\cite{Palladium-cat-rev,Bernardi1997}. The bond-formation process of $d^{10}$-transition metals and olefins can be dissected using the energy decomposition analysis (EDA) and an orbital entanglement
analysis, which allows us to identify the most important orbital interaction along the reaction coordinate.

In the EDA developed by Morokuma~\cite{Morokuma-EDA-1976,Morokuma-EDA-1977}, Ziegler and Rauk~\cite{Bickelhaupt2007,Ziegler1977_EDA,Ziegler1979_EDA,Ziegler1979_2_EDA}, the quantum system is divided into disjoint fragments according to the interaction of interest. In this work, the interaction energy between ligand(s) and the metal center, $\Delta E_{\rm int}$, is decomposed into three main components,
\begin{equation}
\Delta E_{\rm int}=\Delta V_{\rm elstat}+\Delta E_{\rm Pauli}+\Delta E_{\rm oi},
\end{equation}
where $\Delta V_{\rm elstat}$ denotes the electrostatic interaction energy, $\Delta E_{\rm Pauli}$ is the repulsive Pauli interaction, and $\Delta E_{\rm oi}$ denotes the 
orbital interaction between the fragments. The EDA has proven to be a very powerful tool for analysing chemical bonds and orbital interactions in many complex chemical systems, including transition metal complexes~\cite{Hopffgarten2012_EDA,Wolters2013,Wolters2014}.

Quantum information theory allows us to quantify the interaction and correlation of orbitals and orbital pairs~\cite{Boguslawski2014,Rissler2006,entanglement_letter,Boguslawski2013}. The entanglement between one orbital and the orbital bath is measured 
by the von Neumann entropy of the reduced density matrix of the orbital of interest, here referred to as one-orbital reduced density matrix. The eigenvalues of the one-orbital reduced
density matrix $\omega_{\alpha}$ are used to calculate the single-orbital entropy~\cite{legeza_dbss},
\begin{equation}
s(1)_i=-\sum_{\alpha}{\omega_{\alpha,i} \ln \omega_{\alpha,i}}.
\end{equation}
We refer the interested reader to refs~\cite{Boguslawski2013,Boguslawski2014,barcza2014entanglement} for more details on how to calculate orbital-reduced density matrices.
Similarly, the entanglement of two orbitals with the orbital bath is quantified by the two-orbital entropy $s(2)_{i,j}$,
\begin{equation}
s(2)_{i,j}=-\sum_{\alpha}{\omega_{\alpha,i,j}\ln \omega_{\alpha,i,j}},
\end{equation}
where $\omega_{\alpha,i,j}$ are the eigenvalues of the two-orbital reduced density matrix.

The correlation between orbital pair $i$ and $j$ can be measured by the orbital pair mutual information~\cite{legeza_dbss,Legeza2006,Rissler2006},
\begin{equation}
 I_{i|j}=\frac{1}{2}(s(2)_{i,j}-s(1)_i-s(1)_j)(1-\delta_{ij}),
\end{equation}
where $\delta_{ij}$ is the Kronecker delta.

Both $s(1)_i$ and $I_{i|j}$ quantify orbital interactions and can be used to identify different types of electron correlation effects~\cite{entanglement_letter,CUO_DMRG,boguslawski2014chemical}, dissect chemical bonding~\cite{orbitalordering,Boguslawski2013,PCCP_bonding,Ru-entanglement,AP1roG-actinides}, and locate transition state structures in molecular systems~\cite{TTNS-LiF,MIT-Fertita-2014,Duperrouzel2015}.  

\begin{figure*}[t]
\centering
\includegraphics[width=1.0\textwidth]{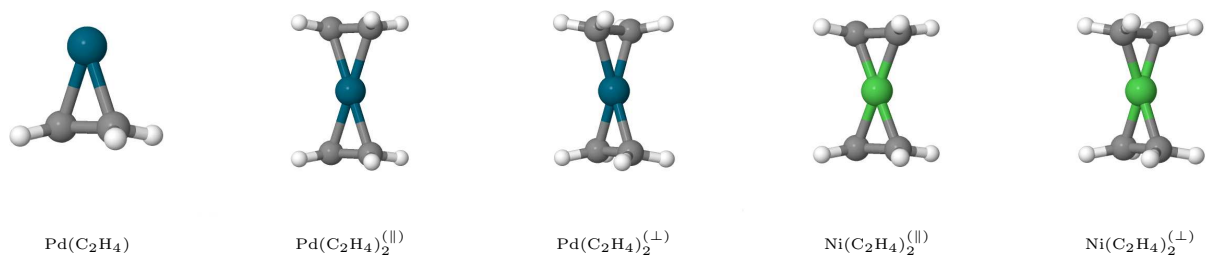}
\caption{Equilibrium structures of nickel- and palladium-ethene complexes optimized by BP86. The $\parallel$ symbol indicates that both ethene ligands are aligned parallel to each other, while the $\perp$ symbol is used to label the perpendicular arrangement of the ethene ligands.}
\label{fig:2}
\end{figure*}

In this work, we investigate the bonding situation in nickel-ethene and palladium-ethene compounds using wavefunction approaches such as the complete active space self-consistent field approach, the density matrix renormalization group algorithm, and coupled cluster theory. In particular, we investigate the potential energy surfaces resulting from the interaction of the ethene molecule(s) approaching the palladium center in three structural rearrangements Pd(C$_2$H$_4$), Pd(C$_2$H$_4$)$_2^{(\parallel)}$, and Pd(C$_2$H$_4$)$_2^{(\perp)}$, where $\parallel$ indicates that the ethene ligands are aligned in parallel, while $\perp$ indicates a perpendicular orientation of the ethene ligands (see Figure~\ref{fig:2}). In case of the nickel-ethene, we augment our previous analysis of the Ni(C$_2$H$_4$) reaction pathway with the symmetric bond formation process of Ni(C$_2$H$_4$)$_2^{(\parallel)}$ and Ni(C$_2$H$_4$)$_2^{(\perp)}$. Furthermore, the bonding interaction between the transition metals (Ni, Pd) and the ethene ligand(s) is analysed in terms of the energy decomposition analysis as implemented in ADF~\cite{Ziegler1977_EDA,Bickelhaupt2007} and an orbital entanglement analysis.

This paper is organized as follows. Section \ref{compu} contains the computational details. In section \ref{result}, the bond formation process of different nickel-ethene and palladium-ethene complexes is 
dissected using the EDA and orbital entanglement analysis. Finally, we conclude in section \ref{concl}.

\begin{figure*}[t]
\centering
\includegraphics[width=1.0\textwidth]{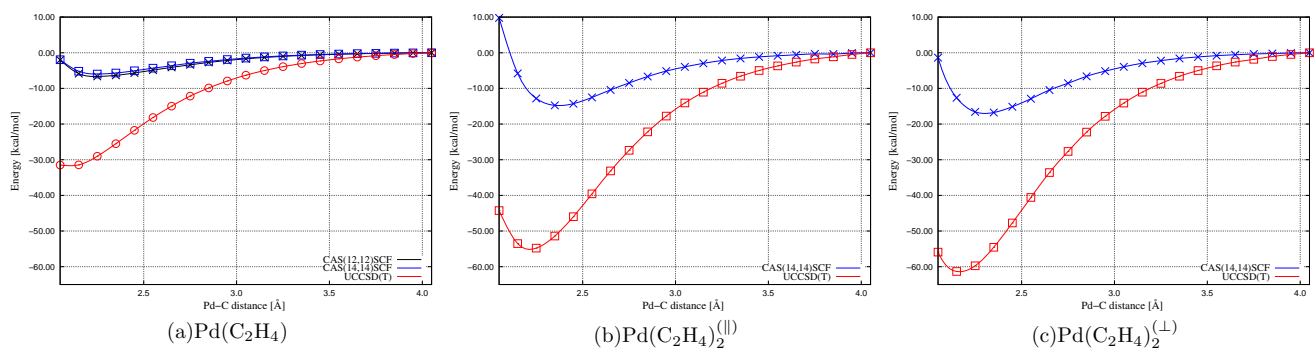}
\caption{Potential energy surfaces for the palladium-ethene reaction pathway in different structural rearrangements. In case of the two ethene molecules, the potential energy surfaces result from the symmetric dissociation of the ethene ligands from the metal centre. The last point of the reaction coordinate is adjusted to zero.}
\label{fig:3}     
\end{figure*}

\section{Computational Details}\label{compu}
\subsection{Geometry Optimization}
The structures of all metal-ethene complexes are optimized by scanning the nickel--carbon bond from 1.75~\AA\ to 2.75~\AA~and the palladium--carbon bond from 2.05~\AA\ to 4.05~\AA\ (constrained geometry optimization). In addition, for $\rm Ni(C_2H_4)_2$, a second reaction pathway was investigated where one ethene molecule approached the $\rm Ni(C_2H_4)$ fragment. In this asymmetric bond formation process, all nickel-carbon distances were fixed, while all hydrogen atoms were allowed to freely relax. 
All calculations were performed with the \textsc{Adf2013} software package~\cite{adf1,adf2,adf2013}. Scalar relativistic effects were included using the ZORA Hamiltonian~\cite{Lenthe1993,VanLenthe1999,VanLenthe1994}. In all calculations, a DZP~\cite{VanLenthe2003} basis set and the BP86~\cite{Perdew86,Becke} exchange--correlation functional were used.

\subsection{CASSCF}
CASSCF~\cite{Roos_casscf,Siegbahn_casscf,Werner_casscf} calculations for nickel-ethene were performed in the \textsc{Dalton2013}~\cite{Dalton2013} software package, while the \textsc{Molpro2012}~\cite{MOLPRO-WIREs,molpro} software suite was used for palladium-ethene. 
A TZP ANO-RCC basis set was employed in all CASSCF calculations with the following contraction schemes: H:$(8s4p3d1f)$ $\rightarrow$ $[6s4p3d1f]$~\cite{TZP_H_He}, C:$(8s7p4d3f2g)\rightarrow[4s3p2d1f]$~\cite{TZP_B_Rn}, \linebreak Ni:$(10s9p8d6f4g2h)\rightarrow[6s5p3d2f1g]$~\cite{TZP_Sc_Hg}, and \linebreak Pd:$(21s18p13d6f4g2h )\rightarrow[10s9p9d6f4g2h]$~\cite{TZP_Sc_Hg}. Scalar relativistic effects were included by the second-order Douglas-Kroll-Hess Hamiltonian~\cite{dkh1,Hess_1986}. The CASSCF orbitals were visualized using the {Jmol14.2.7}\cite{Jmol} visualization software.  	

For Ni(C$_2$H$_4$)$_2^{(\perp)}$ ($n=3$) as well as Pd(C$_2$H$_4$)$_2^{(\perp)}$ and Pd(C$_2$H$_4$)$_2^{(\parallel)}$ ($n=4$), we correlated 14 electrons in 14 orbitals, including the $nd_{xy}$, $nd_{xz}$, $nd_{yz}$, $nd_{z^2}$, $nd_{x^2-y^2}$, and $(n+1)d_{xy}$, $(n+1)d_{xz}$, $(n+1)d_{yz}$, $(n+1)d_{z^2}$, $(n+1)d_{x^2-y^2}$ orbitals from the $d^{10}$-transition metals and both $\pi$ and $\pi^*$ orbitals from the ethene ligands. 
For Ni(C$_2$H$_4$)$_2^{(\perp)}$, the Ni $4d_{xy}$-orbital was excluded resulting in CAS(14,13)SCF calculations. 

For the Pd(C$_2$H$_4$) complex, we performed CAS(12,12)\-SCF and CAS(14,14)SCF calculations. The CAS(12,12) active space contains the metal $4d$ and the ethene $\pi-$ and $\pi^*-$ orbitals. To evaluate the contribution of the $\sigma$ and $\sigma^*$ orbitals, the CAS(12,12) active space was extended to 14 electrons correlated in 14 orbitals in our CAS(14,14)SCF calculations.
The resulting CASSCF orbitals along the potential energy surfaces are presented in Figures S1--S18 of the Supporting Information. 

$\rm C_{2v}$ symmetry was imposed for Pd(C$_2$H$_4$), $\rm D_{2h}$ symmetry for Ni(C$_2$H$_4$)$_2^{(\parallel)}$ and  Pd(C$_2$H$_4$)$_2^{(\parallel)}$, and $\rm D_{2}$ symmetry for Ni(C$_2$H$_4$)$_2^{(\perp)}$ and Pd(C$_2$H$_4$)$_2^{(\perp)}$. 

\subsection{UCCSD and UCCSD(T)}
The Unrestricted Coupled Cluster Singles Doubles (UCCSD) and UCCSD and perturbative Triples (UCCSD(T))~\cite{uccsd-molpro} calculations were performed with the \textsc{Molpro2012}~\cite{MOLPRO-WIREs,molpro} program. The core orbitals were kept frozen, while all virtual orbitals were correlated. The same basis sets, point group symmetries, and relativistic Hamiltonian were used as in our CASSCF calculations. The UCCSD(T) energies are collected in Tables S1-S5 of the Supporting Information.

\subsection{EDA}
The energy decomposition analysis calculations were performed for the nickel- and palladium-ethene complexes at equilibrium distance using the \textsc{ADF2013}~\cite{adf1,adf2,adf2013} software package. Specifically, the supra-molecule was divided into one fragment containing the metal center and a second fragment containing the ethene ligand for monoligated complexes, while a third fragment comprising the second ethene ligand was added for the biligated metal compounds.

\subsection{DMRG}
The \textsc{Budapest DMRG}~\cite{dmrg_ors} program was used to perform the DMRG calculations. As orbital basis, the natural orbitals obtained from the largest CASSCF calculations as described in the previous subsection were used. For the biligated nickel-complexes, the active spaces were extended by including additional occupied and virtual natural orbitals. 10 additional occupied orbitals (2$\times$A$_{\rm g}$, 2$\times$B$_{\rm 3u}$, 1$\times$B$_{\rm 2u}$, 1$\times$B$_{\rm 1g}$, 1$\times$B$_{\rm 1u}$,1$\times$B$_{\rm 2g}$, 1$\times$B$_{\rm 3g}$ and 1$\times$A$_{\rm u}$ for Ni(C$_2$H$_4$)$_2^{(\parallel)}$, and 2$\times$A, 2$\times$B$_{\rm 1}$, 3$\times$B$_{\rm 3}$ and 3$\times$B$_{\rm 2}$ for Ni(C$_2$H$_4$)$_2^{(\perp)}$) and 10 virtual orbitals for Ni(C$_2$H$_4$)$_2^{(\parallel)}$ (2$\times$A$_{\rm g}$, 1$\times$B$_{\rm 3u}$, 1$\times$B$_{\rm 2u}$, 1$\times$B$_{\rm 1g}$, 1$\times$B$_{\rm 1u}$,1$\times$B$_{\rm 2g}$, 2$\times$B$_{\rm 3g}$ and 1$\times$A$_{\rm u}$) and 9 virtual orbitals  for Ni(C$_2$H$_4$)$_2^{(\perp)}$ (4$\times$A, 1$\times$B$_{\rm 1}$, 2$\times$B$_{\rm 3}$ and 2$\times$B$_{\rm 2}$) were added to the active space, increasing it to 34 electrons correlated in 33 orbitals (DMRG(34,33)). The DMRG calculations for Pd complexes were carried out with the same active spaces as in CASSCF. Furthermore, we made sure that the active spaces contained similar orbitals along the reaction coordinate, i.e., molecular orbitals with similar atomic contributions.

To enhance convergence, we optimized the orbital ordering~\cite{orbitalordering}. The initial guess was generated using the dynamically extended-active-space procedure (DEAS)~\cite{legeza_dbss}. In all DMRG calculations, the Davidson diagonalization threshold was set to 10$^{-6}$ for the nickel-complexes, and 10$^{-7}$ for the palladium compounds. The minimum number of block states, $m$, was set to 64 (in the preoptimization), while the maximum number was set to 1024. The convergence of DMRG with respect to $m$ is summarized in Tables S1--S3 of the Supporting Information.

The orbital entanglement and correlations diagrams were determined from the DMRG wavefunctions as described in ref~\cite{Boguslawski2013}.

\begin{figure}[t]
\centering
\includegraphics[width=0.9\columnwidth]{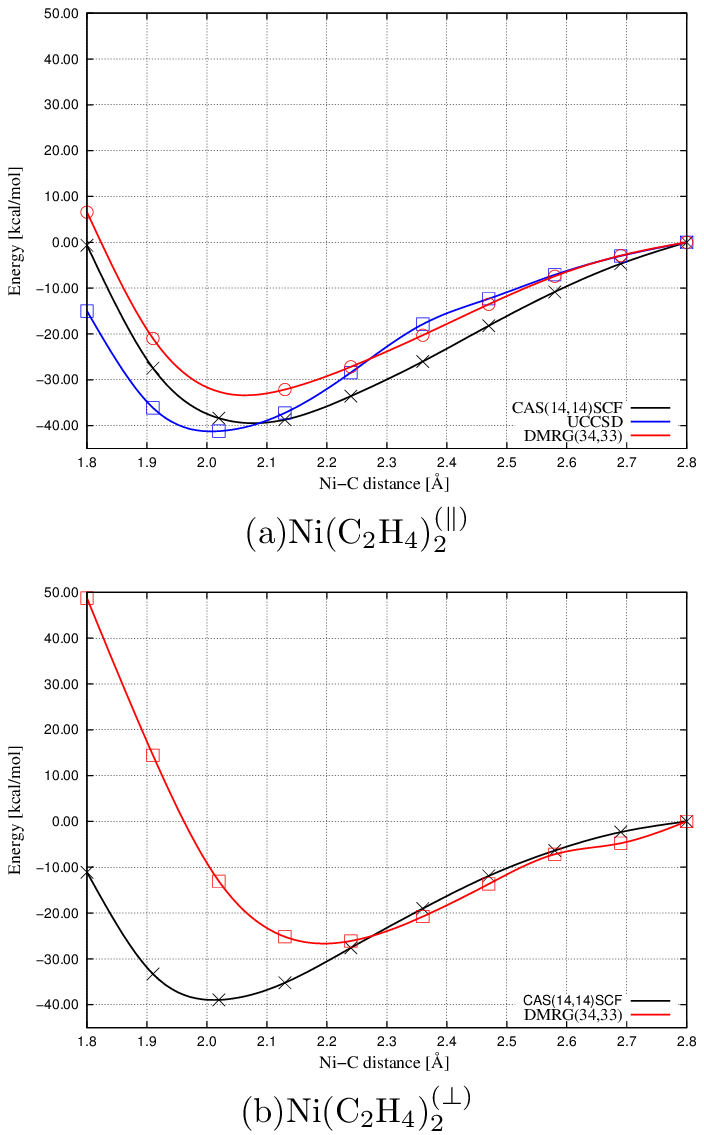}
\caption{Potential energy surfaces for the nickel-ethene reaction pathway in different structural rearrangements. The potential energy surfaces correspond to the symmetric dissociation of the ethene ligands from the metal centre. The last point of the reaction coordinate is adjusted to Zero.}
\label{fig:4}
\end{figure}
\begin{table}[b]
\centering
\caption{Bonding energies for metal-ethene complexes in kcal/mol for CASSCF and CCSD(T).}
\label{tab:1}       
\begin{tabular}{lrrr}
\hline\noalign{\smallskip}
Molecular			& CASSCF	&UCCSD(T)\\
\noalign{\smallskip}\hline\noalign{\smallskip}
Ni(C$_2$H$_4$)$_2^{(\parallel)}$& $>$39.0  &  n/a  \\
Ni(C$_2$H$_4$)$_2^{(\perp)}$ 	& $>$39.0  &   n/a  \\
Pd(C$_2$H$_4$)		& 6.6   & 29.0  \\
Pd(C$_2$H$_4$)$_2^{(\parallel)}$& 14.8  & 51.4  \\
Pd(C$_2$H$_4$)$_2^{(\perp)}$ 	& 16.8  & 54.5  \\
\noalign{\smallskip}\hline
\end{tabular}
\end{table}
\begin{table*}[t]
\centering
\caption{Energy decomposition analysis for nickel- and palladium-ethene complexes. $r_e$ is the equilibrium distance between the metal center and the carbon atom of the ethene molecule(s). $\Delta V_{\rm elstat}$ is the electrostatic interaction energy, $\Delta E_{\rm Pauli}$ is the repulsive Pauli interaction, and $\Delta E_{\rm oi}$ refers to the orbital interaction between the fragments. $E_{\rm int} = \Delta E_{\rm Pauli}+\Delta V_{\rm elstat}+\Delta E_{\rm oi}$ is the total interaction energy.}
\label{tab:2}       
\begin{tabular}{lrrrrr}
\hline\noalign{\smallskip}
Molecule			& $r_e$ [\AA{}] 	& $\Delta E_{\rm Pauli}$ [$\rm\frac{kcal}{mol}$]	& $\Delta V_{\rm elstat}$ [$\rm\frac{kcal}{mol}$]	& $\Delta E_{\rm oi}$ [$\rm\frac{kcal}{mol}$] 	& $E_{\rm int}$ [$\rm\frac{kcal}{mol}$]\\
\noalign{\smallskip}\hline\noalign{\smallskip}
$\rm Ni(C_2H_4)$                 &1.88   &385.3	  &-220.1 (47\%)  &-251.7 (53\%)    &-86.6	\\
$\rm Ni(C_2H_4)_2^{(\parallel)}$&2.02   & 591.9	  &-313.4 (45\%)  &-387.2 (55\%)    &-108.6\\
$\rm Ni(C_2H_4)_2^{(\perp)}$ 	&2.02   & 595.3   &-314.5 (43\%)  &-411.2 (57\%)    &-133.7\\
$\rm Pd(C_2H_4)$                &2.25   &125.2	  &-109.4 (58\%)  &-58.2 (42\%)	    &-42.4\\
$\rm Pd(C_2H_4)_2^{(\parallel)}$&2.35   & 184.4	  &-165.9 (66\%)  &-83.4 (34\%)	    &-64.9\\
$\rm Pd(C_2H_4)_2^{(\perp)}$ 	&2.35   & 184.7   &-166.5 (65\%)  &-89.6 (35\%)	    &-77.3\\
\noalign{\smallskip}\hline
\end{tabular}
\end{table*}
\section{Numerical Results}\label{result}
\subsection{Geometries and Potential Energy Surfaces}

The DFT-optimized geometries along the metal-ethene dissociation pathway are summarized in the Supporting Information. Figure~\ref{fig:2} shows the equilibrium structures of all investigated metal-ethene compounds. For all optimized structures, the hydrogen atoms of the ethene molecule are slightly bent outside the molecular (C--C--H) plane of the uncoordinated ethene.

The potential energy surfaces for the palladium-ethene dissociation process are displayed in Figure~\ref{fig:3}, while Table~\ref{tab:1} summarizes the bonding energy, \emph{i.e.}, the energy difference between the equilibrium structure and the dissociation limit. In general, Pd(C$_2$H$_4$)$_2^{(\perp)}$ has the largest energy to association in both CASSCF and UCC\-SD(T) calculations. The missing dynamic electron correlation energy in CASSCF leads to more shallow potential energy well depths compared to the CC results. In contrast to the monoligated nickel-ethene complex~\cite{Duperrouzel2015}, all investigated quantum chemistry methods predict no barrier to association along the palladium-ethene reaction coordinate.

Figure~\ref{fig:4} shows the potential energy surfaces for the symmetric nickel-ethene dissociation pathway predicted by CASSCF, DMRG, and UCCSD. We were unable to converge the constrained geometry optimization for Ni--C distances larger than 2.8 \AA{}. 
Thus, only estimated potential well depths and bonding energies of Ni(C$_2$H$_4$)$_2^{(\parallel)}$ and Ni(C$_2$H$_4$)$_2^{(\perp)}$ are provided. 
Furthermore, we encountered convergence difficulties in our UCCSD calculations for stretched distances of Ni(C$_2$H$_4$)$_2$. 
Therefore, it remains unclear if the symmetric dissociation of Ni(C$_2$H$_4$)$_2$ features a transition state as found for the monoligated nickel-ethene complex.

As shown in Table~\ref{tab:1}, the lower bound for the bonding energy is considerably larger in nickel-ethene than in palladium-ethene complexes suggesting a stronger bonding interaction between the nickel center and the ethene ligands in terms of $\pi$-donation and metal-to-ligand back-bonding.

In general, CC calculations predict shorter metal-ethene bond lengths than found in CASSCF, which can be attributed to the missing dynamic electron correlation effects in the latter. Specifically, the CASSCF Pd--C equilibrium distance in Pd(C$_2$H$_4$) is approximately 2.25 \AA{}, which reduces to 2.10 \AA{} in UCCSD(T). For Pd(C$_2$H$_4$)$_2^{(\parallel)}$, the equilibrium bond length decreases from 2.35 \AA{} in CASSCF to 2.25 \AA{} in UCCSD(T), while the equilibrium bond lengths are slightly shorter for Pd(C$_2$H$_4$)$_2^{(\perp)}$: 2.35 \AA{} in CASSCF and 2.15 \AA{} in UCCSD(T).

\FloatBarrier
\subsection{Elucidating the Metal-to-Ligand Back-Donation}
\subsubsection{Energy Decomposition Analysis}
The EDA results at the equilibrium distance for all metal-ethene complexes are summarized in Table~\ref{tab:2}. The total interaction energy (see Table~\ref{tab:2}) is defined as the sum of $\Delta E_{\rm Pauli}$, $\Delta V_{\rm elstat}$, and $\Delta E_{\rm oi}$ and quantifies the interaction between the fragments.

All investigated nickel-ethene complexes have a considerably larger total interaction energy (in absolute value) than the corresponding palladium-ethene compounds with similar coordination sphere. Comparing $\Delta V_{\rm elstat}$ with $\Delta E_{\rm oi}$, we observe that $\Delta V_{\rm oi}$ constitutes the dominant contribution in nickel-ethene complexes, while $\Delta V_{\rm elstat}$ dominates in palladium-ethene complexes. Since $\Delta V_{\rm elstat}$ corresponds to the classic electrostatic interaction between fragments and $\Delta E_{\rm oi}$ represents the interaction between orbitals on one fragment with the orbitals on the other fragment, a larger contribution of $\Delta V_{\rm elstat}$ indicates more ionic interactions between the fragments, while a larger contribution of $\Delta E_{\rm oi}$ suggests a stronger covalent nature of the interaction between the fragments. The different ratios between $\Delta V_{\rm elstat}$ and $\Delta E_{\rm oi}$ suggest that the nickel--ligand bond is more covalent, while the palladium--ligand bond is more ionic.

\begin{table}
\centering
\caption{Relation between the strength of orbital entanglement and correlation and electron correlation effects.}
\label{tab:3}  
\begin{tabular}{lll}
\hline\noalign{\smallskip}
Correlation effects & $s(1)_i$ & $I_{i|j}$\\
\noalign{\smallskip}\hline\noalign{\smallskip}
Nondynamic &  $>$0.5 & $\approx10^{-1}$\\
Static & 0.5-0.1 & $\approx10^{-2}$\\
Dynamic & $<$0.1 & $\approx10^{-3}$\\
\noalign{\smallskip}\hline
\end{tabular}
\end{table}

\begin{figure*}[t]
\centering
\includegraphics[width=1.9\columnwidth]{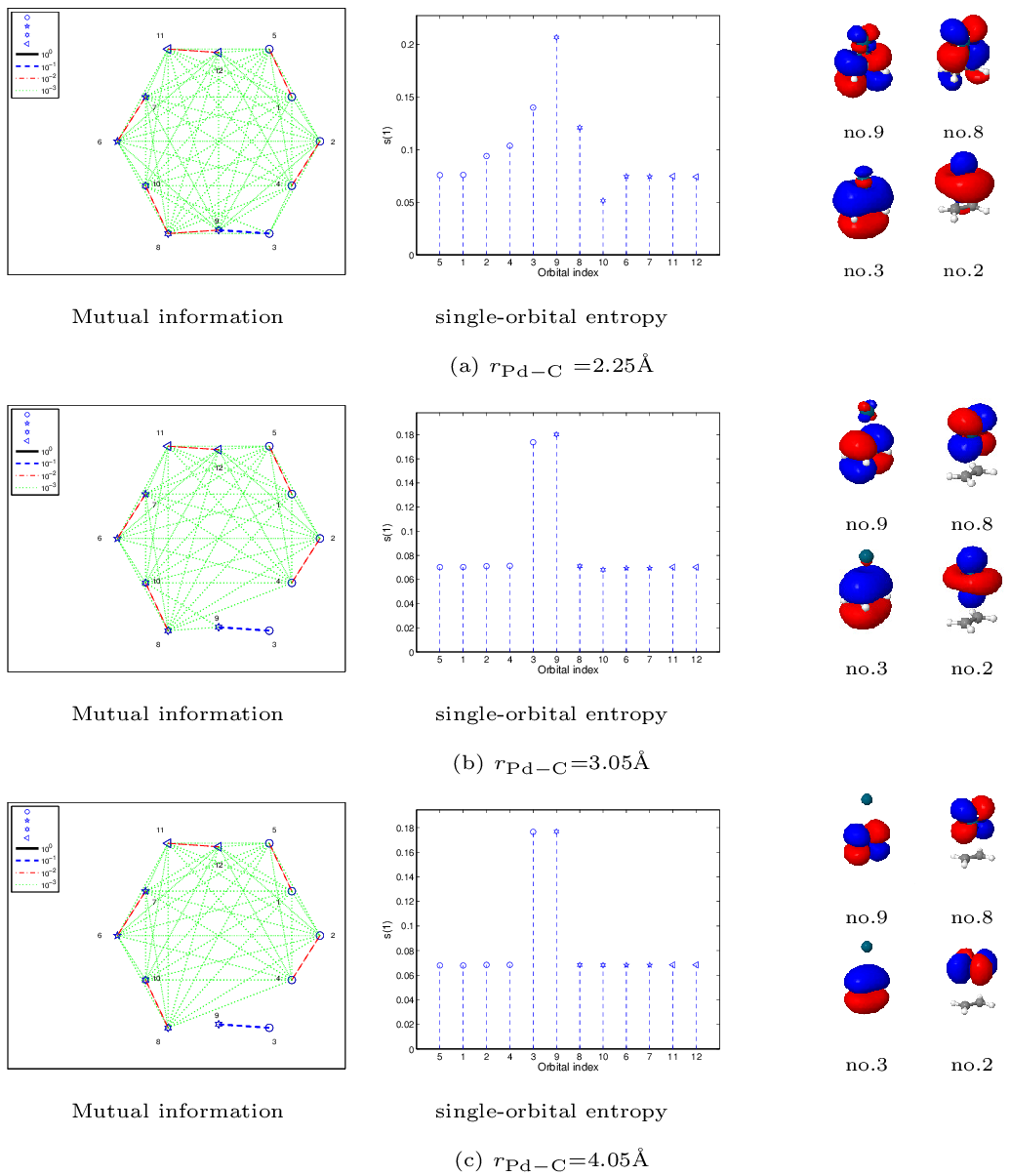}
\caption{Orbital-pair mutual information and single-orbital entropy for Pd(C$_2$H$_4$) determined from DMRG(14,14) calculations.}
\label{fig:5}     
\end{figure*}

\begin{figure*}
\centering
\includegraphics[width=0.9\textwidth]{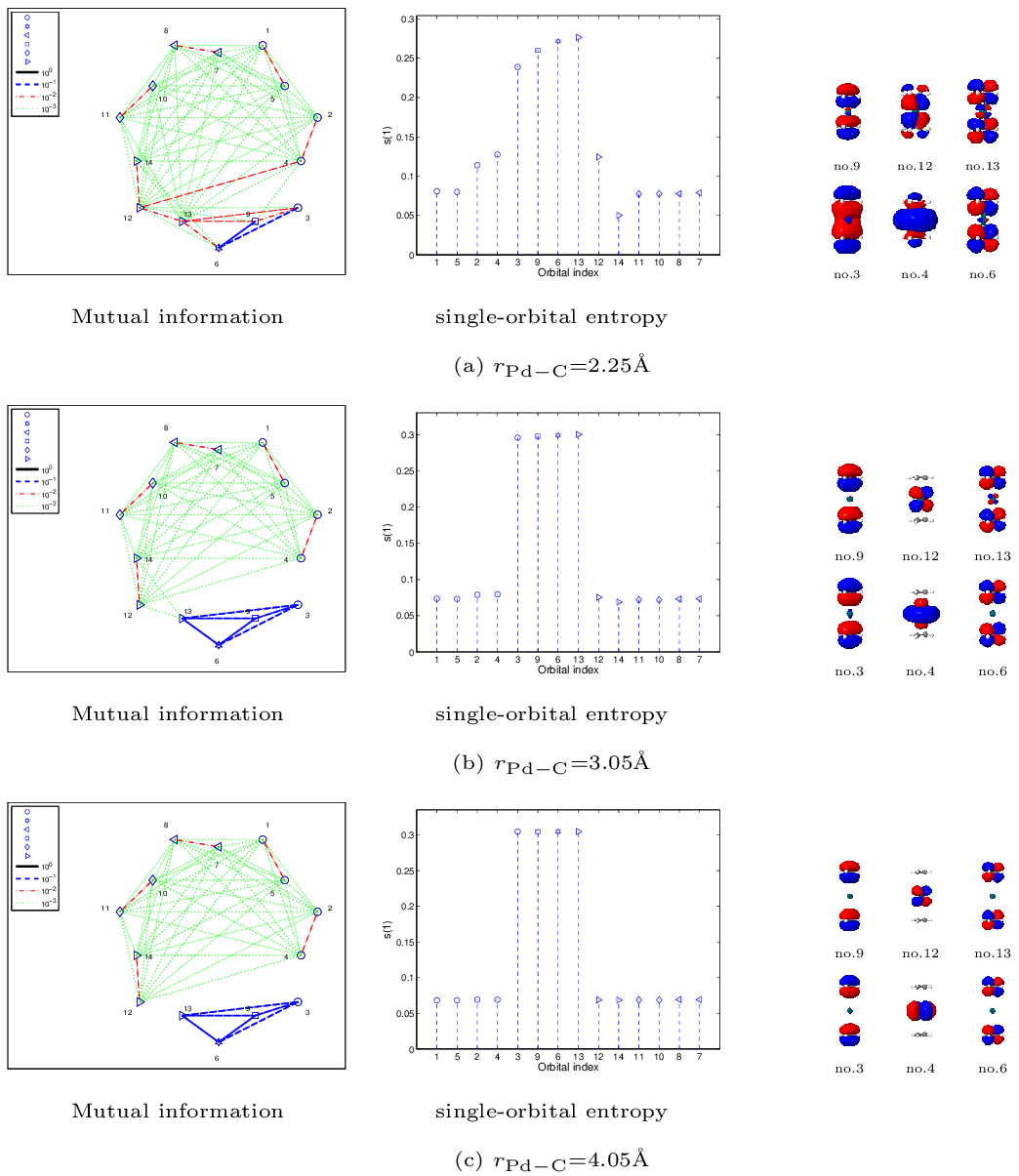}
\caption{Orbital-pair mutual information and single-orbital entropy for Pd(C$_2$H$_4$)$_2^{\parallel}$ determined from DMRG(14,14) calculations.}
\label{fig:6}     
\end{figure*}
\begin{figure*}
\centering
\includegraphics[width=0.9\textwidth]{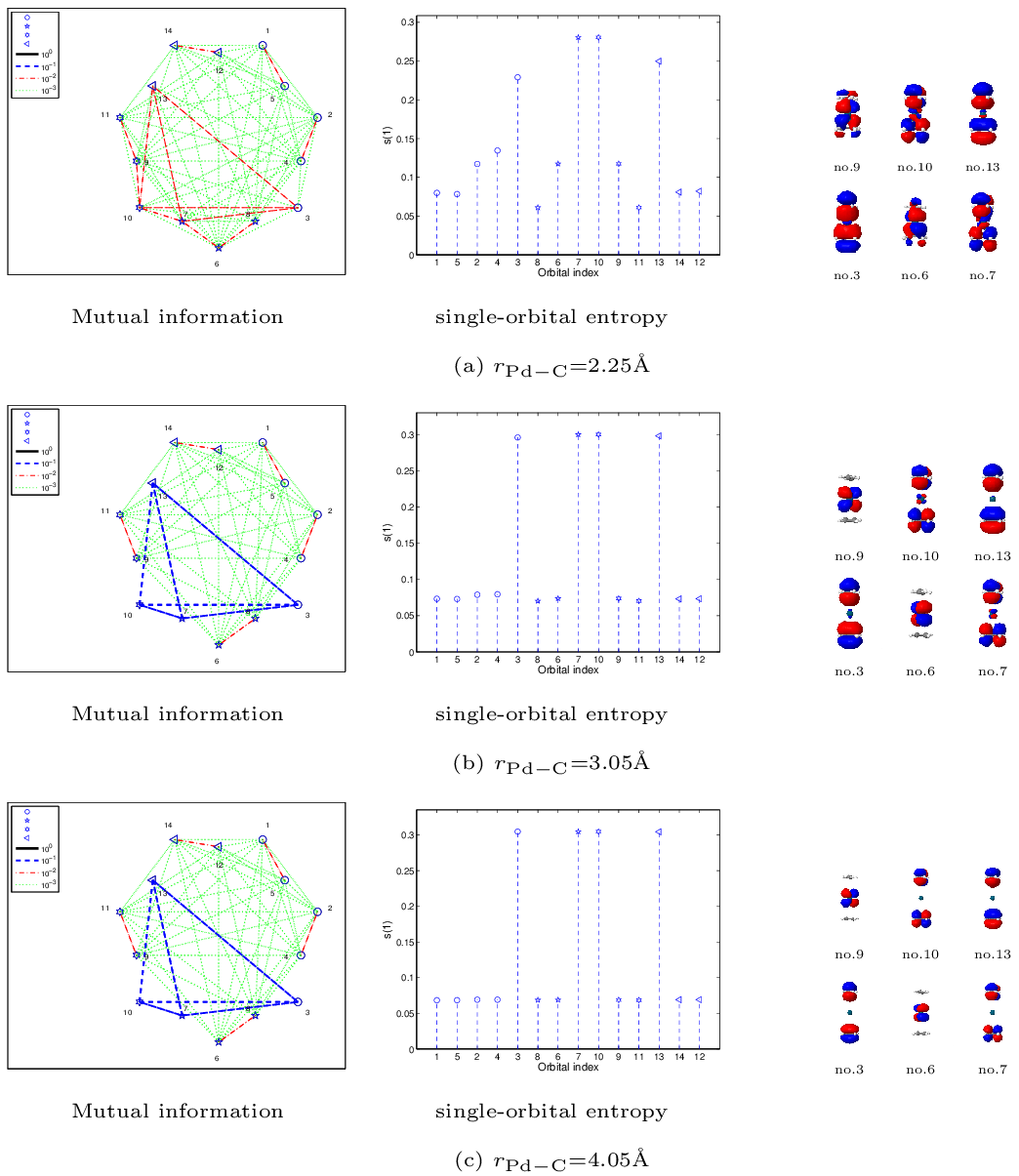}
\caption{Orbital-pair mutual information and single-orbital entropy for Pd(C$_2$H$_4$)$_2^{\perp}$ determined from DMRG(14,14) calculations.}
\label{fig:7}     
\end{figure*}

\begin{figure*}[t]
\centering
\includegraphics[width=0.9\textwidth]{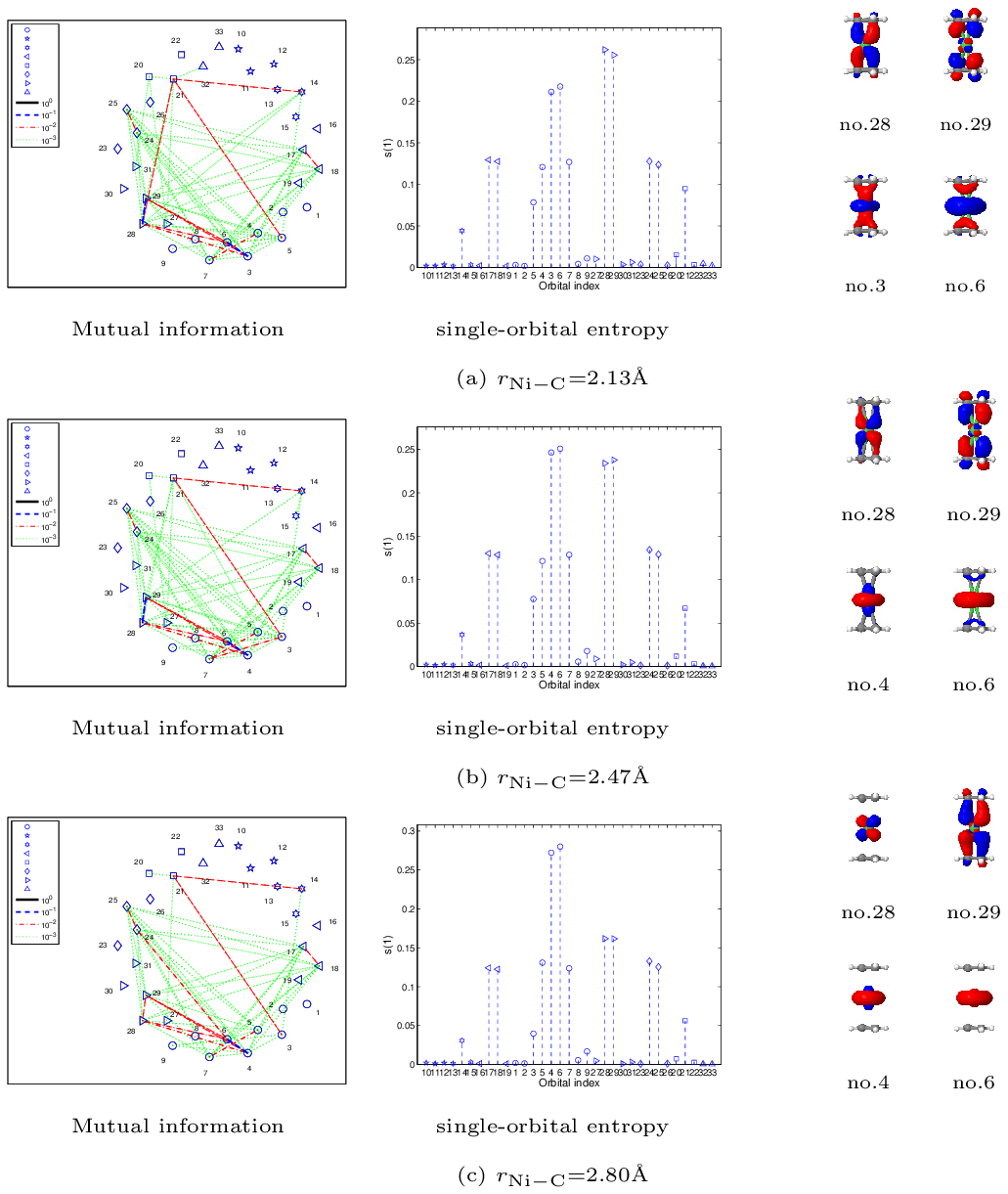}
\caption{Orbital-pair mutual information and single-orbital entropy for Ni(C$_2$H$_4$)$_2^{\parallel}$ determined from DMRG(34,33) calculations.}
\label{fig:8}     
\end{figure*}

\begin{figure*}[t]
\centering
\includegraphics[width=0.9\textwidth]{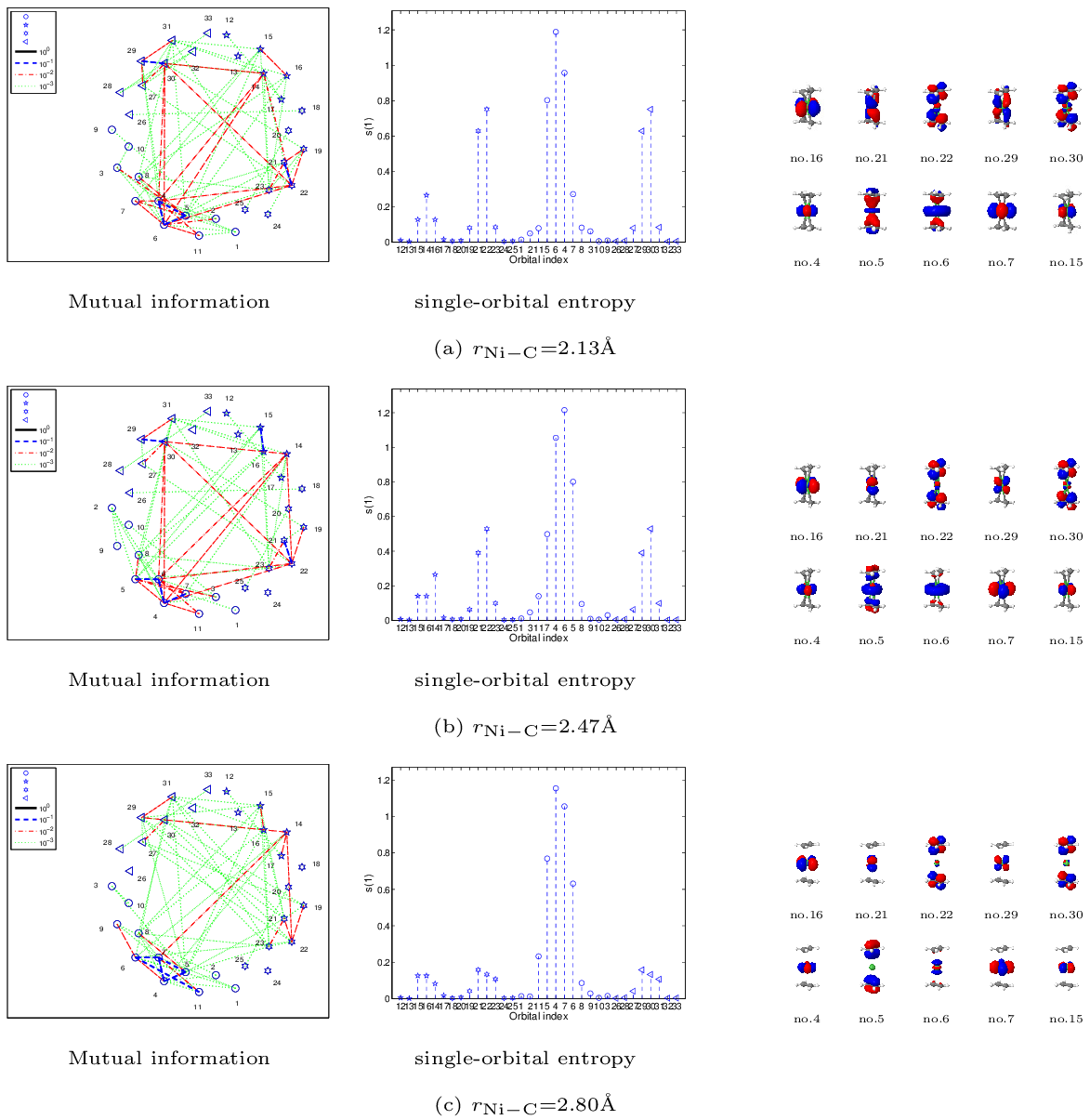}
\caption{Orbital-pair mutual information and single-orbital entropy for Ni(C$_2$H$_4$)$_2^{\perp}$ determined from DMRG(34,33) calculations.}
\label{fig:9}     
\end{figure*}

\subsubsection{Orbital Entanglement}
An orbital entanglement analysis uses the single-orbital entropy $s(1)_i$ to measure orbital entanglement and the orbital-pair mutual information $I_{i|j}$ to quantify the correlation between orbital pairs. Both $s(1)_i$ and $I_{i|j}$ are represented using diagrams. Specifically, the strength of the orbital-pair mutual information is colour-coded. Strongly correlated orbital pairs are connected by blue lines ($I_{i|j} \approx 10^{-1}$), moderately correlated orbitals by red lines ($I_{i|j} \approx 10^{-2}$), while weakly correlated orbitals are indicated by green lines ($I_{i|j} \approx 10^{-3}$), etc. As presented in ref~\cite{entanglement_letter}, the strength of orbital entanglement and correlation can be associated with electron correlation effects~\cite{Bartlett1994,Bartlett2007} (see Table~\ref{tab:3}).
Since we are interested in bond formation processes, our analysis will focus on orbitals and orbital pairs with moderately to strongly entangled orbitals, \emph{i.e.}, orbitals with $s(1)_i > 0.1$ and $I_{i|j}>10^{-2}$.

\paragraph{Orbital Entanglement and Correlation in Palladium-Ethene}

Figures~\ref{fig:5}-\ref{fig:7} show the mutual information and single-orbital entropy for palladium-ethene complexes at different points of the reaction coordinate (see Figure~\ref{fig:3}) along with the strongly entangled molecular orbitals.

For the monoligated Pd(C$_2$H$_4$) in the dissociation limit (Figure~\ref{fig:5}(c)), molecular orbitals centered on the metal atom (Pd $4d$- and $5d$-orbitals) and on the ethene fragment ($\pi$- and $\pi^*$-orbitals) are correlated. No significant orbital correlations can be observed between orbitals centered on different fragments. The most strongly correlated orbitals are the ethene $\pi$- and $\pi^*$-orbitals (nos.~3 and 9 in Figure~\ref{fig:5}(c)). When the ethene molecule approaches the metal center (see Figure~\ref{fig:5}(b)), the Pd $d_{yz}$-orbital (no.~8) and the ethene $\pi^*$-orbital (no.~9) become weakly correlated. This orbital correlation corresponds to the metal-to-ligand \linebreak back-donation process. However, the most strongly correlated orbitals remain centered on the ethene ligand ($\pi$-$\pi^*$) and the metal atom ($4d$-$5d$), respectively. Around the equilibrium structure, the molecular orbitals involved in metal-to-ligand back-bonding (nos.~8 and 9 in Figure~\ref{fig:5}(a)) become moderately correlated, while molecular orbitals involved in $\pi$-donation from the ethene $\pi$-orbitals to the Pd $4d_{z^2}$-orbital (nos.~2 and 3 in Figure~\ref{fig:5}(a)) are only weakly correlated. The dominant orbital correlations remain between the ethene $\pi$- and $\pi^*$-orbitals and between Pd $4d$- and $5d$-orbitals.

A similar trend in the orbital correlation and entanglement diagrams can be observed for Pd(C$_2$H$_4$)$_2^{(\parallel)}$ (see Figure~\ref{fig:6}). In the dissociation limit (Figure~\ref{fig:6}(c)), orbital correlations remain distributed among the ethene $\pi$- and $\pi^*$-orbitals (nos.~3, 6, 9, and 13) as well as Pd $4d$- and $5d$-orbitals (nos.~1, 2, 4, 5, 7, 8, 10, 11, 12, 14). When both ethene ligands approach the Pd center (see Figure~\ref{fig:6}(b)), the Pd $d_{yz}$-orbital (no.~12) and the ligand $(\pi^*_1+\pi^*_2)$-orbital (no.~13) are weakly correlated. These orbitals are involved in the metal-to-ligand back-bonding process. However, the dominant orbital correlations remain centered on the ligand orbitals and on the metal orbitals, respectively. Around the equilibrium structure, the changes in the correlation and entanglement patterns are more profound than for the monoligated palladium-ethene complex. While the Pd $d_{yz}$-orbital (no.~12) and the ligand $(\pi_1+\pi_2)$-orbital (no.~13) are moderately correlated, the correlation strength between the ligand $(\pi^*_1+\pi^*_2)$-orbital and the remaining bonding and antibonding combinations of the $\pi$- and $\pi^*$-orbitals decreases. Similarly, the Pd $d_{z^2}$-orbital and the ligand $(\pi_1+\pi_2)$-orbital are weakly entangled,1 suggesting a negligible contribution of $\pi$-donation in the bond-formation process of Pd(C$_2$H$_4$)$_2^{(\parallel)}$.

For Pd(C$_2$H$_4$)$_2^{(\perp)}$, the Pd $d$-orbitals and the ligand $\pi$- and $\pi^*$-orbitals remain uncorrelated in the dissociation limit and for stretched palladium-ethene distances (see Figure~\ref{fig:7}(b) and (c)). In contrast to Pd(C$_2$H$_4$) and Pd(C$_2$H$_4$)$_2^{(\parallel)}$, the correlation and entanglement diagrams drastically change around the equilibrium distance (see Figure~\ref{fig:7}(a)). At this point, the Pd $4d_{yz}$-orbital (no.~9) and the ligand $(\pi^*_1,\pi^*_2)$-orbital (no.~10) as well as the Pd $4d_{xz}$-orbital (no.~6) and the ligand $(\pi^*_3,\pi^*_4)$-orbital (no.~7) are moderately correlated. Furthermore, the correlation between the ligand $\pi$ and $\pi^*$-orbitals (nos.~3, 7, 10, and 13) reduces and all ligand orbitals are only moderately correlated compared to the ligand orbitals in Pd(C$_2$H$_4$) and Pd(C$_2$H$_4$)$_2^{(\parallel)}$. These dominant correlations between metal and ligand orbitals suggest that the electronic structure of Pd(C$_2$H$_4$)$_2^{(\perp)}$ features two metal-to-ligand back-bonding interactions. This \emph{double-back-bonding} mechanism may lead to an additional stabilization of the Pd(C$_2$H$_4$)$_2^{(\perp)}$ isomer compared to the Pd(C$_2$H$_4$)$_2^{(\parallel)}$ complex and elucidates the larger orbital interaction energy $\Delta E_{\rm oi}$ of Pd(C$_2$H$_4$)$_2^{(\perp)}$ in the EDA.

\paragraph{Orbital Entanglement and Correlation in Nickel-Ethene}

The entanglement and correlation diagrams for Ni(C$_2$H$_4$)$_2^{(\parallel)}$ at different points along the reaction coordinate are shown in Figure~\ref{fig:8}. In the dissociation limit, the leading orbital correlation is found between the Ni $3d_{z^2}$- and $4s$-orbitals (nos.~4 and 6 in Figure~\ref{fig:8}(c)). In contrast to Pd(C$_2$H$_4$)$_2^{(\parallel)}$, the metal $3d_{yz}$ and ligand $(\pi^*_1+\pi^*_2)$-orbitals are already moderately correlated. Their correlation further increases when the ethene ligands approach the metal center (see Figure~\ref{fig:8}(b)). Around the equilibrium structure, the strong correlation between the Ni $3d_{z^2}$- and $4s$-orbitals diminishes and the bonding and antibonding combination of the Ni $3d_{z^2}$-orbital and the ligand $(\pi^*_1+\pi^*_2)$-orbital (nos.~3 and 6 in Figure~\ref{fig:8}(a)) are strongly correlated. The latter orbital correlation corresponds to the metal-to-ligand $\pi$-donation mechanism. In contrast to the monoligated Ni(C$_2$H$_4$) complex (see ref~\cite{Duperrouzel2015} for details), $\pi$-donation does not commence until close to the equilibrium geometry and the corresponding orbital correlations are comparable to those between the Ni $3d_{yz}$ and ligand $(\pi^*_1+\pi^*_2)$-orbitals.

The reaction pathway of Ni(C$_2$H$_4$)$_2^{(\perp)}$ features a notably different evolution of orbital correlation and entanglement compared to its structural isomer Ni(C$_2$H$_4$)$_2^{(\parallel)}$. In the vicinity of dissociation (Figure~\ref{fig:9}(c)), the Ni $3d_{x^2-y^2}$- and $3d_{z^2}$-orbitals (nos.~4, 6, and 7) as well as the ligand $(\pi_1,\pi_2)^*$-orbital (no.~5) are strongly correlated with each other, while the orbitals involved in metal-to-ligand back-bonding (nos.~21 and 22 as well as nos.~29 and 30) are moderately correlated. When the ethene ligands approach the metal center (Figure~\ref{fig:9}(b)), metal and ligand orbitals that participate in $\pi$-donation (nos.~5 and 6) and metal-to-ligand back-bonding (nos.~21 and 22 as well as nos.~29 and 30) are strongly correlated. Close to the equilibrium geometry (Figure~\ref{fig:9}(a)), we observe a transition of orbital entanglement and correlation patters. Specifically, the correlation between the Ni $3d$- and $4d$-orbitals (nos.~15 and 16 as well as nos.~4 and 7) decreases when approaching the equilibrium geometry. Similar to Pd(C$_2$H$_4$)$_2^{\perp}$, the orbital correlation and entanglement analysis suggest that Ni(C$_2$H$_4$)$_2^{(\perp)}$ features two metal-to-ligand back-bonding interactions which may stabilize the Ni(C$_2$H$_4$)$_2^{(\perp)}$ complex compared to the Ni(C$_2$H$_4$)$_2^{(\parallel)}$ isomer.

\paragraph{Comparison of the Bonding Mechanism in Ni- and Pd-Olefines}

Finally, we will compare the bond formation process and the bonding interactions in nickel-ethene and palladium-ethene complexes. Both the EDA and orbital entanglement analysis highlight the different nature of the metal-ethene bond and of the bond-formation mechanism. In general, the bonding interaction in the nickel-ethene complexes is stronger than in the corresponding palladium-ethene compounds. Furthermore, the degree of covalency of the metal-olefin bond is higher for nickel-ethene than for palladium-ethene (\emph{cf.} the large values of $\Delta E_{\rm oi}$ for Ni(C$_2$H$_4$)$_x$ compared to Pd(C$_2$H$_4$)$_x$ ($x=1,2$). The different bonding nature and bond-formation processes is supported by our orbital entanglement analysis. While the metal-to-ligand back-bonding mechanism plays an important role in the bond-formation process in nickel-olefin compounds, and which establishes for stretched Ni--C$_2$H$_4$ distances in the vicinity of dissociation, the metal-to-ligand back-bonding in Pd(C$_2$H$_4$)$_x$ becomes important close to the equilibrium geometry. Similarly, the role of $\pi$-donation considerably differs in nickel- and palladium-ethene. Specifically, our entanglement analysis predicts that the $\pi$-donation mechanism is insignificant in Pd(C$_2$H$_4$)$_x$ complexes, while it forms an essential part in the bond-formation process in Ni(C$_2$H$_4$)$_x$ compounds where the correlation between the ligand $\pi$-orbitals and the metal $d_{z^2}$-orbital increases when the ethene ligands approach the nickel center.

\FloatBarrier

\section{Conclusions}\label{concl}

In this work, we investigated the interactions between ethene ligands and the nickel and palladium center along the metal-ethene reaction coordinate and for perpendicular and parallel orientations of the ethene ligands. 
While both nickel and palladium are $d^{10}$-transition metals, they exhibit a considerably distinct bonding mechanism and interactions with ethene ligands. 
Specifically, nickel--carbon bonds are shorter and stronger than palladium--carbon bonds for both the parallel and perpendicular orientation. Moreover, the bond between nickel and ethene has predominantly covalent character, while the palladium--ethene bond has mainly ionic character. 

Both $d^{10}$-transition metals create more stable complexes with ethene in perpendicular orientation, where two metal-to-ligand back-bonding mechanisms can be observed. The double-back-bonding allows for stronger orbital interactions. Moreover, our entanglement analysis indicates that molecular orbitals involved in $\pi$-donation from the ethene $\pi$-orbitals to the metal $d_{z^2}$ -orbital are considerably more correlated in nickel-ethene than palladium-ethene complexes. Thus, while $\pi$-donation plays an important role in the bond-formation process of nickel-ethene, the palladium--ethene bond does not feature strong $\pi$-donation.

This work demonstrates that concepts from quantum information theory constitute a useful and complementary tool to well-established methods like energy decomposition analysis in dissecting chemical reactions.  
 
\begin{acknowledgements}
We gratefully acknowledge financial support from the Natural Sciences and Engineering Research Council of Canada (NSERC) and the Hungarian Research Fund (OTKA K100908 and NN110360). 
K.B.~acknowledges financial support from the Swiss National Science Foundation (P2EZP2 148650) and the Banting Postdoctoral Fellowship program.
C.D.~acknowledges financial support from the McMaster Chemistry \& Chemical Biology Summer Research Scholarship and the NSERC Undergraduate Student Research Award fellowship.

The authors acknowledge support for computational resources from \textsc{SHARCNET}, a partner consortium in the Compute Canada national HPC platform. 
\end{acknowledgements}

\end{document}